\documentclass[prl,twocolumn,showpacs,preprintnumbers,amsmath,amssymb,citeautoscript]{revtex4-1}
 
\usepackage{graphicx}
\usepackage{amssymb, amsfonts, amsmath}
\usepackage{dcolumn}
\usepackage{bm}
\usepackage{color}

\newlength{\figwidth}
\begin{document}

\title{Giant superconducting fluctuations in the compensated semimetal FeSe at the BCS-BEC crossover}

\author{S.\;Kasahara$^1$}
\author{T.\;Yamashita$^1$}
\author{A.\;Shi$^1$}
\author{R.\;Kobayashi$^2$}
\author{Y.\;Shimoyama$^1$}
\author{T.\;Watashige$^1$}
\author{K.\;Ishida$^1$}
\author{T.\;Terashima$^2$}
\author{T.\;Wolf$^3$}
\author{F.\;Hardy$^3$}
\author{C.\;Meingast$^3$}
\author{H.\;v.\;L\"ohneysen$^3$}
\author{A.\;Levchenko$^4$}
\author{T.\;Shibauchi$^{5,*}$}
\author{Y.\;Matsuda$^{1,\dagger}$}

\affiliation{
$^1$Department of Physics, Kyoto University, Kyoto 606-8502, Japan\\
$^2$Research Center for Low Temperature and Materials Sciences, Kyoto University, Kyoto 606-8501, Japan\\
$^3$Institute of Solid State Physics, Karlsruhe Institute of Technology, D-76021 Karlsruhe, Germany\\
$^4$Department of Physics, University of Wisconsin-Madison, Madison, Wisconsin 53706, USA\\
$^5$Department of Advanced Materials Science, University of Tokyo, Kashiwa, Chiba 277-8561, Japan\\
$^*$e-mail: {\sf shibauchi@k.u-tokyo.ac.jp}
$^\dagger$e-mail: {\sf matsuda@scphys.kyoto-u.ac.jp}
}




\maketitle

{\bf
The physics of the crossover between weak-coupling Bardeen-Cooper-Schrieffer (BCS) and strong-coupling Bose-Einstein-condensate (BEC) limits gives a unified framework of quantum bound (superfluid) states of interacting fermions. This crossover has been studied in the ultracold atomic systems \cite{SadeMelo2008,Randeria2014,Nasc2010,Sagi2014}, but is extremely difficult to be realized for electrons in solids. Recently, the superconducting semimetal FeSe with a transition temperature $T_{\rm c}=8.5$\,K has been found to be deep inside the BCS-BEC crossover regime \cite{Kasahara2014,Terashima2014,Watson2015}. Here we report experimental signatures of preformed Cooper pairing in FeSe below $T^*\sim20$\,K, whose energy scale is comparable to the Fermi energies. In stark contrast to usual superconductors, large nonlinear diamagnetism by far exceeding the standard Gaussian superconducting fluctuations is observed below $T^*\sim20$\,K, providing thermodynamic evidence for prevailing phase fluctuations of superconductivity. Nuclear magnetic resonance (NMR) and transport data give evidence of pseudogap formation at $\sim T^*$. The multiband superconductivity along with electron-hole compensation in FeSe may highlight a novel aspect of the BCS-BEC crossover physics.
}

In the BCS regime weakly coupled pairs of fermions form the condensate wave function, while in the BEC regime the attraction is so strong that the fermions form local molecular pairs with bosonic character.  The physics of the crossover is described by two length scales, the average pair size or coherence length $\xi_{\rm pair}$ and the average interparticle distance $1/k_{\rm F}$, where $k_{\rm F}$ is the Fermi wave number.  In the BCS regime the pair size is very large and $k_{\rm F} \xi_{\rm pair }\gg 1$, while local molecular pairs in the BEC regime lead to $k_{\rm F}\xi_{\rm pair}\ll1$.   The crossover regime is characterised by $k_{\rm F}\xi_{\rm pair}\sim 1$, or equivalently the ratio of superconducting gap to Fermi energy $\Delta/\varepsilon_{\rm F}$ of the order of unity. In this crossover regime, the pairs interact most strongly and new states of interacting fermions may appear; preformed Cooper pairing at much higher temperature than $T_{\rm c}$ is theoretically proposed \cite{SadeMelo2008,Randeria2014}. Experimentally, however, such preformed pairing associated with the BCS-BEC crossover has been controversially debated  in ultracold atoms \cite{Nasc2010,Sagi2014} and cuprate superconductors \cite{Emery1994,Corson1999,Li2010,Keimer2015}. Of particular interest is the pseudogap formation associated with the preformed pairs that lead to a suppression of low-energy single-particle excitations. Also important is the breakdown of Landau's Fermi-liquid theory due to the strong interaction between fermions and fluctuating bosons. 
In ultracold-atomic systems, this crossover has been realised by tuning the strength of the interparticle interaction via the Feshbach resonance. In these artificial systems, Fermi-liquid-like behaviour has been reported in thermodynamics even in the middle of crossover \cite{Nasc2010}, but more recent photoemission experiments have suggested a sizeable pseudogap opening and a breakdown of the Fermi-liquid description \cite{Sagi2014}. 

On the other hand, for electron systems in bulk condensed matter it has been extremely difficult to access the crossover regime.  Perhaps the most frequently studied systems have been underdoped high-$T_{\rm c}$ cuprate superconductors \cite{Emery1994,Corson1999,Li2010,Keimer2015} with substantially shorter coherence length than conventional superconductors.  In underdoped cuprates, pseudogap formation and non-Fermi-liquid behaviour are well established, and unusual superconducting fluctuations have also been found above $T_{\rm c}$ \cite{Corson1999,Li2010}. However, the pseudogap appears at a much higher temperature than the onset temperature of superconducting fluctuations \cite{Keimer2015}. It is still unclear whether the system is deep inside the crossover regime and to what extent the crossover physics is relevant to the phase diagram in underdoped cuprates. 
It has been also suggested that in iron-pnictide BaFe$_2$(As$_{1-x}$P$_x$)$_2$ the system may approach the crossover regime in the very vicinity of a quantum critical point \cite{Hashimoto2012,Shibauchi2014}, but the fine-tuning of the material to a quantum critical point by chemical substitution is hard to accomplish.  
Therefore this situation calls for a search of new systems in the crossover regime.

\begin{figure}[t]
\includegraphics[width=1.0\linewidth]{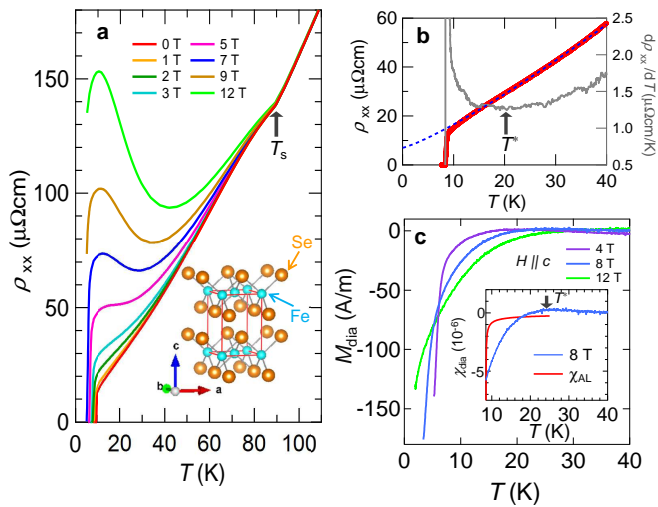}
\caption{{\bf Excess conductivity and diamagnetic response of a high-quality single crystal of FeSe.} {\bf a,}  $T$-dependence of  $\rho_{xx}$ in magnetic fields ($\bm{H} \parallel c$).  The structural transition occurs at $T_{\rm s}=90$\,K.  Inset shows the crystal structure of FeSe.  {\bf b,} $T$-dependence of $\rho_{xx}$ (red) and $d\rho_{xx}/dT$ (gray).  Below $T^*$ shown by arrow, $\rho_{xx}$ shows  a downward curvature.  The blue dashed line represents $\rho_{xx}(T)=\rho_0+AT^{\alpha}$ with $\rho_0=7$\,$\mu \Omega$cm $A$=0.6\,$\mu \Omega$ cm/K$^2$ and $\alpha=1.2$.   {\bf c,} Diamagnetic response in magnetization $M_{\rm dia}$ for $\bm{H} \parallel c$.  The inset shows the diamagnetic susceptibility $\chi_{\rm dia}$ at 8\,T (blue) compared with the estimated $\chi_{\rm AL}$ in the standard Gaussian fluctuations theory (red). 
 } 
\end{figure}

Among different families of iron-based superconductors, iron chalcogenides FeSe$_x$Te$_{1-x}$ exhibit the strongest band renormalisation due to electron correlations, and recent angle-resolved photoemission spectroscopy (ARPES)  studies for $x=0.35-0.4$ have shown that some of the bands near the Brillouin zone centre have very small Fermi energy, implying that the superconducting electrons in these bands are in the crossover regime \cite{Lubashevsky2012,Okazaki2014}. 
Among the members of the iron chalcogenide series, FeSe ($x=0$) with the simple crystal structure formed of tetrahedrally bonded layers of iron and selenium (Fig.\,1a, inset), is particularly intriguing.
FeSe undergoes a tetragonal-orthorhombic structural transition at $T_{\rm s}\approx90$\,K which is accompanied by a kink in the temperature dependence of in-plane resistivity $\rho_{xx}(T)$ (Fig.\,1a), but in contrast to other Fe-based superconductors, no long-range magnetic ordering occurs at any temperature. Recently, the availability of high-quality bulk single crystals grown by chemical vapor transport \cite{Boehmer2013} has reopened investigations into the electronic properties of FeSe.   Several experiments performed on these crystals have shown that all Fermi-surface bands are very shallow \cite{Kasahara2014,Terashima2014,Watson2015};  one or two electron pockets centred at the Brillouin-zone corner with Fermi energy $\varepsilon_{\rm F}^e\sim 3$\,meV, and a compensating cylindrical hole pocket near the zone centre with  $\varepsilon_{\rm F}^h\sim 10$\,meV.   FeSe is a multigap superconductor with two distinct superconducting gaps $\Delta_1\approx$ 3.5\,meV and $\Delta_2\approx$ 2.5\,meV \cite{Kasahara2014}.  Remarkably, the Fermi energies are comparable to the superconducting gaps; $\Delta/\varepsilon_{\rm F}$ is $\sim0.3$ and $\sim1$ for hole and electron bands, respectively \cite{Kasahara2014}.  These large $\Delta/\varepsilon_{\rm F}(\approx \frac{1}{k_{\rm F}\xi_{\rm pair}})$ values indicate that FeSe is in the BCS-BEC crossover regime.  In fact, values of  $2\Delta_1/k_BT_{\rm c}\approx 9$ and $2\Delta_2/k_BT_{\rm c}\approx 6.5$, which are significantly enhanced with respect to the weak-coupling BCS value of 3.5, imply that the attractive interaction holding together the superconducting electron pairs takes on an extremely strong-coupling nature, as expected in the crossover regime.   Moreover,  the appearance of a new high-field superconducting phase  when the Zeeman energy is comparable to the gap and  Fermi energies, $\mu_0H\sim \Delta \sim \varepsilon_{\rm F}$, suggests a peculiar superconducting state of FeSe \cite{Kasahara2014}.     Therefore FeSe provides a new platform to study the electronic properties in the crossover regime.

It is well known that thermally fluctuating droplets of Cooper pairs can survive above $T_{\rm c}$. These fluctuations arise from amplitude fluctuations of the superconducting order parameter and have been investigated for many decades.  Their effect on thermodynamic, transport, and thermoelectric quantities in most superconductors is well understood in terms of standard Gaussian fluctuation theories \cite{Book}.  However, in the presence of preformed pairs associated with the BCS-BEC crossover, superconducting fluctuations are expected to be strikingly enhanced compared to Gaussian theories due to additional phase fluctuations.  Moreover, it has been suggested that such enhanced fluctuations can lead to a reduction of the density of states (DOS), dubbed the pseudogap \cite{SadeMelo2008,Randeria2014}. 

Quite generally, superconducting fluctuations give rise to an enhancement of the normal-state conductivity, which manifests itself as a downturn towards lower $T$ of the resistivity vs. temperature curve above $T_{\rm c}$.  
The high-field magnetoresistance of compensated semimetals is essentially determined by the product of the  scattering times of electron and hole bands \cite{Kasahara2014}. The large, insulating-like upturn in $\rho_{xx}(T)$ at high fields is thus an indication of the high quality of our crystals (Fig.\,1a). At low temperatures, however, the expected downturn behaviour is observed, implying large superconducting fluctuations. Even at zero field, 
${\rm d}\rho_{xx}(T)/{\rm d}T$ shows a minimum around $T^* \sim 20$\,K (Fig.\,1b), indicating the appearance of excess conductivity below  $\sim T^*$. However, a quantitative analysis of this excess conductivity is difficult to achieve because it strongly depends on the extrapolation of the normal-state resistivity above $T^*$ to lower $T$.   In addition, the resistivity may be affected by a change of the scattering time when a pseudogap opens at $T^*$ as observed in underdoped cuprates \cite{Kontani2008}.

\begin{figure*}[t]
\includegraphics[width=0.85\linewidth]{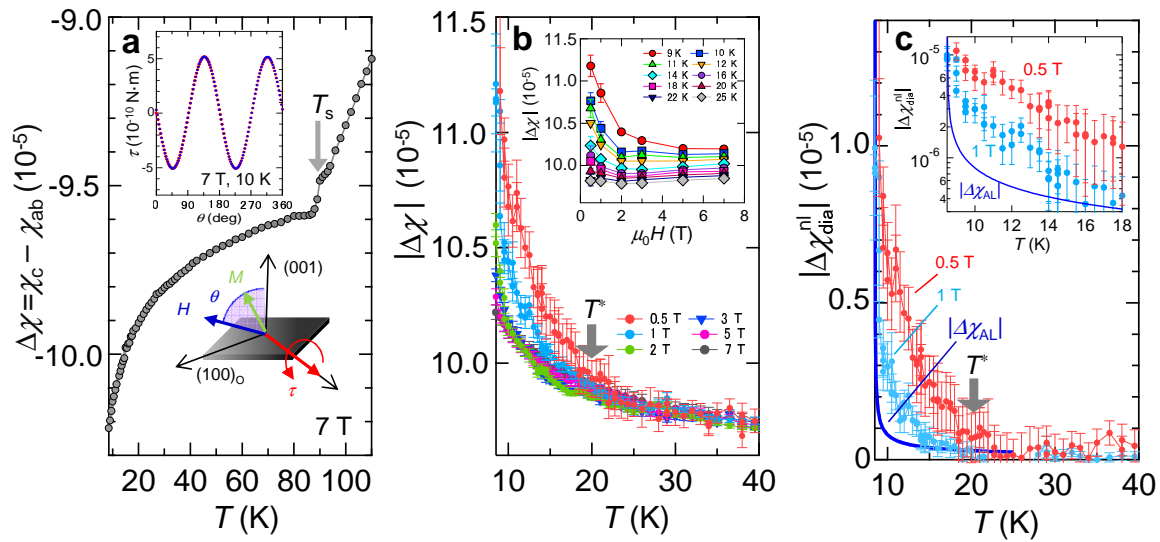}
\caption{{\bf Diamagnetic response detected by magnetic torque measurements above $T_{\rm c}$.} {\bf a,} Anisotropy of the susceptibility between the $c$ axis and $ab$ plane, $\Delta \chi$, at 7\,T.   The lower inset is schematics of the  $\theta$-scan measurements.   The upper inset shows the  magnetic torque $\tau$ as a function of $\theta$.   Torque curves measured by rotating {\boldmath $H$} in clockwise (red) and anticlockwise (blue) directions coincide (the hysteresis component is less than 0.01\% of the total torque).  {\bf b,}  The main panel shows the $T$-dependence of $|\Delta \chi|$  at various magnetic fields.     Inset shows the $H$-dependence of $|\Delta \chi|$ at fixed temperatures.  {\bf c,} Temperature dependence of the non-linear diamagnetic response at $\mu_0H=0.5$\,T (red) and 1\,T (blue) obtained by $|\Delta \chi_{\rm dia}^{\rm nl}| \approx \Delta \chi(H)-\Delta \chi (7 \rm~T)$ .  Blue line represents the estimated $|\Delta \chi_{\rm AL}|$ in the standard Gaussian fluctuations theory.  The inset displays $|\Delta \chi_{\rm dia}^{\rm nl}|$  plotted in a semi-log scale at low temperatures.
} 
\end{figure*}

We therefore examine the superconducting fluctuations in FeSe through the diamagnetic response in the magnetization.  The magnetization $M(T)$ for magnetic field $\bm{H}$ parallel to the $c$ axis (Fig.\,S1) exhibits a downward curvature below $\sim T^*$.  This pronounced decrease of $M(T)$ can be attributed to the diamagnetic response due to superconducting fluctuations.  Figure\,1c shows the diamagnetic response in the magnetization $M_{\rm dia}$ between 0 and 40\,K  for $\mu_0H=4,8$ and 12\,T, obtained by subtracting a constant $M$ as determined at 30\,K.  
Although there is some ambiguity due to weakly temperature dependent normal-state susceptibility, we find a rough crossing point in $M_{\rm dia}(T,H)$ near $T_{\rm c}$. Such a crossing behaviour is considered as a typical signature of large fluctuations, and has been found in cuprates \cite{Welp1991}. 
The thermodynamic quantities do not include the Maki-Thompson type fluctuations. Hence the fluctuation-induced diamagnetic susceptibility of most superconductors including multiband systems can be well described by the standard Gaussian-type (Aslamasov-Larkin; AL) fluctuation susceptibility
$\chi_{\rm AL}$ \cite{Ullah1991,Ussishkin2002,NbSe2}, which is given by 
\begin{equation}
\chi_{\rm AL} \approx -\frac{2\pi^2}{3}\frac{k_{\rm B}T_{\rm c}}{\Phi_0^2}\frac{\xi_{ab}^2}{\xi_c}\sqrt{\frac{T_{\rm c}}{T-T_{\rm c}}}
\end{equation}
 in the zero-field limit  \cite{AL1975}. Here $\Phi_0$ is the flux quantum, and $\xi_{ab}$ ($\xi_c$) is the effective coherence lengths parallel (perpendicular) to the $ab$ plane at zero temperature.  In the multiband case,  the behaviour of $\chi_{\rm AL}$ is determined by the shortest coherence length of the main band, which governs the orbital upper critical field. The diamagnetic contribution $\chi_{\rm AL}$ is expected to become smaller in magnitude at higher fields, and thus $|\chi_{\rm AL}|$ yields an upper bound for the standard Gaussian-type amplitude fluctuations. 
In the inset of Fig.\,1c, we compare $\chi_{\rm dia}$ at 8\,T with $\chi_{\rm AL}$, where we use $\xi_{ab}$ ($=5.5$\,nm) and $\xi_c$ ($=1.5$\,nm)  \cite{Kasahara2014,Terashima2014}. Obviously the amplitude of $\chi_{\rm dia}$ of FeSe is much larger than that expected in the standard theory, implying that the superconducting fluctuations in FeSe are distinctly different from those in conventional superconductors.

The highly unusual nature of superconducting fluctuations in FeSe can also be seen in the low-field diamagnetic response.  Since the low-field magnetization below 2\,T is not reliably obtained from conventional magnetization measurements, we resort to sensitive torque magnetometry.   The magnetic torque $\bm{\tau}=\mu_0 V\bm{M}\times \bm{H}$ is a thermodynamic quantity which has a high sensitivity for detecting magnetic anisotropy.  Here $V$ is the sample volume, $\bm{M}$ is the induced magnetization and $\bm{H}$ is the external magnetic field.  For our purposes, the most important advantage of this method is that an isotropic Curie contribution from impurity spins is cancelled out \cite{Watanabe2012}. 
At each temperature and field, the angle-dependent torque curve $\tau(\theta)$ is measured in $\bm{H}$ rotating within the $ac$ $(bc)$ plane, where $\theta$ is the polar angle from the $c$ axis (Fig.\,2a, lower inset).   In this geometry, the difference between the $c$-axis and $ab$-plane susceptibilities, $\Delta\chi =\chi_c-\chi_{ab}$,  yields a $\pi$-periodic oscillation term with respect to $\theta$ rotation, $\tau_{2\theta}(T,H,\theta)=\frac{1}{2}\mu_0H^2V \Delta\chi  (T,H)\sin2\theta$ (Fig.\,2a, upper inset and Fig.\,S2) \cite{Okazaki2011,Kasahara2012}.  In the whole measurement range, $\Delta\chi$ is negative, i.\,e., $\chi_a>\chi_c$, which is consistent with magnetic-susceptibility \cite{Grechnev2013} and NMR Knight-shift measurements \cite{Baek2015,Boehmer2015}.  Figure\,2a shows the $T$ dependence of $\Delta\chi$ at 7\,T, which is determined by the amplitude of the sinusoidal curve.  At $T_{\rm s}$, $\Delta\chi(T)$ exhibits a clear anomaly associated with the tetragonal-orthorhombic structural transition. Upon approaching $T_{\rm c}$, $\Delta\chi$ shows a diverging behaviour.   Figure\,2b and its inset depict the $T$ and $H$ dependence of $|\Delta \chi|(T,H)$, respectively.  Above $T^*\sim20$\,K, $|\Delta \chi|(T,H)$ is nearly field-independent.   Below $T^*$, however, $|\Delta \chi|(T,H)$ increases with decreasing $H$,  indicating nonlinear $H$ dependence of $M$.    This nonlinearity increases steeply with decreasing temperature. Since $|\Delta \chi|$ points to a diverging behaviour in the zero-field limit upon approaching $T_{\rm c}$ (Fig.\,2b, inset),  this strongly non-linear behaviour is clearly caused by superconducting fluctuations.

Thus the diamagnetic response of FeSe contains $H$-linear and nonlinear contributions to the magnetization; $\Delta \chi (T)$ can be written as  $\Delta \chi =\Delta \chi_{\rm dia}^{\rm nl}+\Delta \chi_{\rm dia}^{\rm l}+\Delta\chi_{\rm N}$, where $\Delta \chi_{\rm dia}^{\rm nl}$ and $\Delta \chi_{\rm dia}^{\rm l}$ represent the diamagnetic contributions from nonlinear and linear field dependence  of magnetization, respectively,  and  $\Delta\chi_{\rm N}$ is the anisotropic part of the normal-state susceptibility, which is independent of $H$.   Since  $\Delta \chi(T)$ is almost  $H$-independent at high fields (inset, Fig.\,2b),  $\Delta\chi_{\rm dia}^{\rm nl}$ is estimated by subtracting $H$-independent terms  from $\Delta \chi$.  In Fig.\,2c, we plot $|\Delta \chi_{\rm dia}^{\rm nl}|$ estimated from  $\Delta \chi_{\rm dia}^{\rm nl}(H) \approx \Delta \chi(H)-\Delta \chi (7\,{\rm T})$, which we compare with the expectation from the Gaussian fluctuation theory at zero field given by
$
\Delta \chi_{\rm AL} \approx -\frac{2\pi^2}{3}\frac{k_{\rm B}T_{\rm c}}{\Phi_0^2}\left(\frac{\xi_{ab}^2}{\xi_c}-\xi_c\right)\sqrt{\frac{T_{\rm c}}{T-T_{\rm c}}}. 
$   
Near $T_{\rm c}$, $\Delta \chi_{\rm dia}^{\rm nl}$ at 0.5\,T is nearly ten times larger than  $\Delta \chi_{\rm AL}$.  It should be noted that since  $|\Delta \chi_{\rm dia}^{\rm nl}|$  increases with decreasing $H$,  $|\Delta \chi_{\rm dia}^{\rm nl}|$ in the zero-field limit should be much larger than  $|\Delta \chi_{\rm dia}^{\rm nl}|$ at 0.5\,T.   Thus the nonlinear diamagnetic response by far dominates the superconducting fluctuations when approaching $T_{\rm c}$ in the zero-field limit.   We note that although the AL diamagnetic contribution contains a nonlinear term visible at low fields, this term is always smaller than the AL fluctuation contribution at zero field \cite{Ullah1991,Ussishkin2002,NbSe2}.

Our magnetization and torque results provide thermodynamic evidence of giant superconducting fluctuations in the normal state of FeSe by far exceeding the Gaussian fluctuations.  We stress that since the energy scale of $k_{\rm B}T^* \sim 2$\,meV is comparable to $\varepsilon_{\rm F}^e$, it is natural to attribute the observed fluctuations to preformed pairs associated with the BCS-BEC crossover.  In the presence of those pairs, superconducting phase fluctuations \cite{Emery1994} arising from the mode coupling of fluctuations are expected to be significantly enhanced and to produce a highly nonlinear diamagnetic response, as observed in the experiments.   This nonlinear response with large amplitude is profoundly different from the Gaussian behaviour in conventional superconductors.

\begin{figure}[t]
\includegraphics[width=0.9\linewidth]{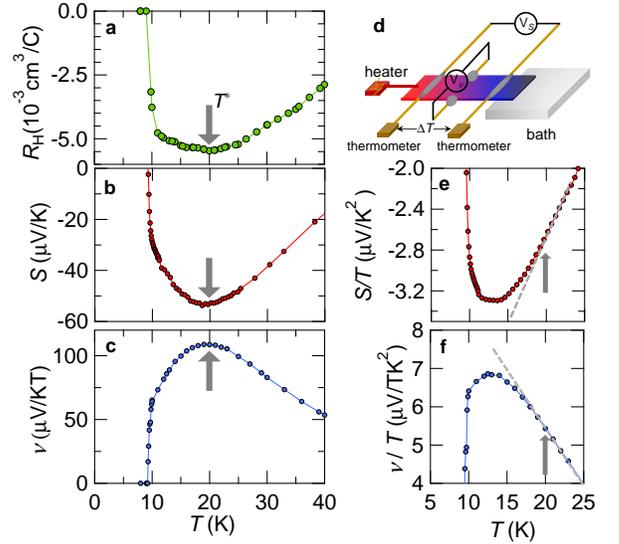}
\caption{ {\bf Possible pseudogap formation below $T^*$ evidenced by NMR and transport measurements.} {\bf a,} Temperature dependence of the NMR relaxation rate divided by temperature $1/T_1T$. Inset: At 14.5\,T, the temperature dependence of $1/T_1T$ between $\sim10$ and 70\,K is fitted to a Curie-Weiss law $\propto (T+16\,{\rm K})^{-1}$ (dashed line). Main panel: The difference between the Curie-Weiss fit and the low field data $\Delta(1/T_1T)$ is plotted as a function of temperature. {\bf b,} Hall coefficient, $R_{\rm H}$. {\bf c,} Seebeck coefficient, $S$. {\bf d,} Nernst coefficient, $\nu$ in the zero-field limit as functions of temperature.  Inset in {\bf d,} is a schematic of the measurement set-up of the  thermoelectric coefficients (see Supplementary Information). 
} 
\end{figure}

Next we discuss the possible pseudogap formation associated with the preformed pairs, which suppresses the DOS and hence leads to a change in quasiparticle scattering. We have measured the relaxation time $T_1$ of $^{77}$Se NMR spectroscopy in FeSe single crystals at different fields applied along the $c$ axis. At 14.5\,T close to the upper critical field, the temperature dependence of $1/T_1T$, which is dominated by the dynamical spin susceptibility $\chi({\bf q})$ at the antiferromagnetic wave vector ${\bf q}=(\pi,\pi)$, can be fitted well by a Curie-Weiss law in a wide temperature range below $T_{\rm s}$ (Fig.\,3a, inset). At low fields of 1 and 2\,T, however, $1/T_1T(T)$ shows a noticeable deviation from this fit (dashed line in Fig.\,3a, inset), and the difference between the fit and the low-field data $\Delta(1/T_1T)$ starts to grow at $\sim T^*$ (Fig.\,3a, main panel). As the superconducting diamagnetism is an orbital effect which is dominated at ${\bf q} =0$, the spin susceptibility $\chi(\pi,\pi)$ is not influenced by the orbital diamagnetism. Therefore, the observed deviation of $1/T_1T(T)$ is a strong indication of a depletion of the DOS, providing spectroscopic evidence for the psedugap formation below $\sim T^*$. The onset temperature and the field dependence of the nonlinear contribution of $1/T_1T(T)$ bear a certain similarity to the features of the diamagnetic susceptibility, pointing to the intimate relation between the pseudogap and preformed pairs in this system. 

The pseudogap formation is further corroborated by the measurements of Hall ($R_{\rm H}$), Seebeck ($S$) and Nernst ($\nu$) coefficients (Figs.\,3b-d).   The negative sign of the Hall and Seebeck data indicates that the transport properties are governed mainly by the electron band, which is consistent with the previous analysis of the electronic structure in the orthorhombic phase below $T_{\rm s}$ \cite{Watson2015}. Obviously, at $T^* \sim 20$\,K all the coefficients show a minimum or maximum. Since the Hall effect is insensitive to superconducting fluctuations, the minimum of $R_{\rm H} (\propto \frac{\sigma_h-\sigma_e}{\sigma_h+\sigma_e})$, where $\sigma_{e(h)}$ is the Hall conductivity of electrons (holes), suggests a change of the carrier mobility at $\sim T^*$.  The thermomagnetic Nernst coefficient consists of two contributions generated by different mechanisms: $\nu=\nu_{\rm N}+\nu_{\rm S}$. The first term represents the contribution of normal quasiparticles. The second term, which is always positive, represents the contribution of fluctuations of either amplitude or phase of the superconducting order parameter.  Upon approaching $T_{\rm c}$, $\nu_{\rm S}$ is expected to diverge \cite{Yamashita2015}.  As shown in Fig.\,3d, however, such a divergent behaviour is absent.  This is because in the present very clean system, $\nu_{\rm N}$ is much larger than $\nu_{\rm S}$ (Fig.\,S4a).  Since $\nu_{\rm N}$ and $S$ are proportional to the energy derivatives of the Hall angle and conductivity at the Fermi level, respectively,  $\nu_{\rm N} \propto (\partial \tan \theta_H/\partial \varepsilon) _{\varepsilon=\varepsilon_{\rm F}}$ and $S\propto (\partial \ln \sigma/\partial \varepsilon)_{\varepsilon=\varepsilon_{\rm F}}$ both sensitively detect the change of the energy dependence and/or anisotropy of the scattering time at the Fermi surface  (see also Figs.\,S4b,c for $\nu/T(T)$ and $S/T(T)$).  Therefore the temperature dependence of the three transport coefficients most likely implies a change in the quasiparticle excitations at $T^*$, which is consistent with the pseudogap formation. 
We also note that anomalies at similar temperatures have been reported for the temperature dependence of the thermal expansion \cite{Boehmer2013} as well as of Young's modulus \cite{Boehmer2015}. Recent scanning-tunnelling-spectroscopy data also suggests some suppression of the DOS at low energies in a similar temperature range \cite{Rossler2015}. 

Let us comment on another thermodynamic quantity, the electronic specific heat, which is also related to the DOS of quasiparticles. The specific heat $C$ at such comparatively high temperatures, however, is dominated by the phonon contribution $\propto T^3$ \cite{Boehmer2015,Lin2011}, which makes it difficult to resolve the pseudogap anomaly. Also the reduction of $C/T$ may partly be cancelled by the increase by the strong superconducting fluctuations found by the present study.  It should be also stressed that FeSe exhibits a semimetallic electronic structure with the compensation condition, i.e. the electron and hole carrier densities should be the identical. 
Such a compensated situation of the electronic structure may alter significantly the chemical potential shift expected in the BEC theories for a single-band electronic structure.  How the entropy in crossover semimetals behaves below $T^*$ is a fundamentally new problem, which deserves further theoretical studies. 

\begin{figure}[t]
\includegraphics[width=0.8\linewidth]{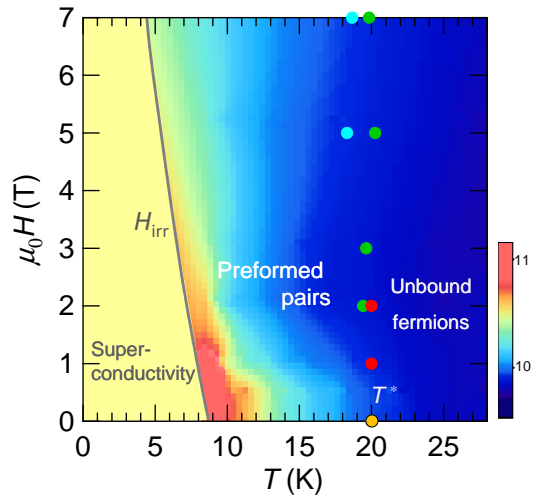}
\caption{{\bf $\bm{H}$-$\bm{T}$ phase diagram of FeSe for $\bm{H}\parallel c$.}  Solid line is the irreversibility line $H_{irr}(T)$ \cite{Kasahara2014}.  The colour represents the magnitude of $\Delta\chi$ (in $10^{-5}$, scale shown in the colour bar) from magnetic torque measurements (Fig.\,2b). Preformed pair regime is determined by the minimum of $d\rho_{xx}(H)/dT$ (blue circles), the peak of Nernst coefficient $\nu_{\rm peak}$ (green circles) and the onset of $\Delta(1/T_1T)$ in the NMR measurements (red circles).  
} 
\end{figure}

Figure\,4 displays the schematic $H$-$T$ phase diagram of FeSe for $\bm{H}\parallel c$.  The fluctuation regime associated with preformed pairing is determined by the temperatures at which $d\rho_{xx}(H)/dT$ shows a minimum (Fig.\;S5a) and $\nu(H)$ shows a peak (Fig.\;S5b) in magnetic fields, as well as by the onset of $\Delta(1/T_1T)$ (Fig.\;3a).  
The diamagnetic signal, NMR relaxation rate, and transport data consistently indicate that the preformed pair regime extends over a wide range of the phase diagram. The phase fluctuations dominate at low fields where the non-linear diamagnetic response is observed (see the inset of Fig.\,2b). This phase-fluctuation region continuously connects to the vortex-liquid regime above the irreversibility field $H_{\rm irr}$, where a finite resistivity is observed with a broad superconducting transition (Fig.\,1a).

Finally, we remark that the preformed Cooper pairs and pseudogap develop in the non-Fermi-liquid state characterized by a linear-in-temperature resistivity, highlighting the highly unusual normal state of FeSe in the BCS-BEC crossover regime.
The resistivity above $T^*$  can be fitted up to $\sim50$\,K as $\rho_{xx}(T)=\rho_{xx}(0)+AT^{\alpha}$ with $\alpha=1.1-1.2$ where the uncertainty arises from the fact that $\rho_{xx}(0)$ is unknown (Fig.\,1b).  Thus the exponent $\alpha$ close to unity indicates a striking deviation from the Fermi-liquid behaviour of $\alpha=2$.   This non-Fermi-liquid behaviour in FeSe is reminiscent of the anomalous normal-state properties of high-$T_{\rm c}$ cuprate superconductors.   The main difference between these systems and FeSe is the multiband nature of the latter;  the Fermi surface consists of compensating electron and hole pockets.  The present observation of preformed pairs together with the breakdown of Fermi-liquid theory in FeSe implies an inherent mechanism that brings about singular inelastic scattering properties of strongly interacting fermions in the BCS-BEC crossover.

\vspace{6mm}

{\bf Acknowledgements} 

We thank K. Behnia, I. Danshita,  H. Kontani, A. Perali, M. Randeria, and Y. Yanase for fruitful discussions. This work was supported by Grants-in-Aid for Scientific Research (KAKENHI) from Japan Society for the Promotion of Science (JSPS), and by the `Topological Material Science' Grant-in-Aid for Scientific Research on Innovative Areas from the Ministry of Education, Culture, Sports, Science and Technology (MEXT) of Japan. The work of A.\,L. was supported by NSF Grant No. DMR-1401908.




\newpage
\begin{center}
{\large \bf Supplementary Information: \\
Giant superconducting fluctuations in the compensated semimetal FeSe at the BCS-BEC crossover}

\bigskip
{S.\;Kasahara$^1$},
{T.\;Yamashita$^1$},
{A.\;Shi$^1$},
{R.\;Kobayashi$^{1,2}$},
{Y.\;Shimoyama$^1$},
{T.\;Watashige$^1$},
{K.\;Ishida$^1$},
{T.\;Terashima$^2$},
{T.\;Wolf$^3$},
{F.\;Hardy$^3$},
{C.\;Meingast$^3$},
{H.\;v.\;L\"ohneysen$^3$},
{A.\;Levchenko$^4$},
{T.\;Shibauchi$^{5,*}$}, and
{Y.\;Matsuda$^{1,\dagger}$}
\\

\medskip
{\it \small
\setlength{\baselineskip}{16pt}
$^1$Department of Physics, Kyoto University, Kyoto 606-8502, Japan\\
$^2$Research Center for Low Temperature and Materials Sciences, Kyoto University, Kyoto 606-8501, Japan\\
$^3$ Institute of Solid State Physics, Karlsruhe Institute of Technology, D-76021 Karlsruhe, Germany\\
$^4$ Department of Physics, University of Wisconsin-Madison, Madison, Wisconsin 53706, USA\\
$^5$ Department of Advanced Materials Science, University of Tokyo, Kashiwa, Chiba 277-8561, Japan\\
$^*$e-mail: {\sf shibauchi@k.u-tokyo.ac.jp}
$^\dagger$e-mail: {\sf matsuda@scphys.kyoto-u.ac.jp}
}

\end{center} 

\bigskip




\renewcommand{\figurename}{{\bf FIG.\,S}$\!\!$}
  \renewcommand{\thefigure}{{\bf\arabic{figure}}}
\renewcommand{\thesection}{SUPPLEMENTARY NOTE\;\arabic{section}}
\renewcommand{\tablename}{\bf Supplementary Table $\!\!$}
\renewcommand{\theequation}{S\arabic{equation}}

\section{Sample characterization}
High-quality single crystals of tetragonal $\beta$-FeSe were grown by low-temperature vapor transport method at  Karlsruhe Institute of Technology and Kyoto University \cite{Boehmer2013}. 
As shown in Fig.\,1b in the main text, taking $\rho_{xx}(T_c^+) \approx 10\,\mu \Omega$\,cm as an upper limit of the residual resistivity, leads to the residual resistivity ratio $RRR>40$. The large $RRR$ value, large magnetoresistance below $T_s$, quantum oscillations at high fields \cite{Terashima2014,Watson2015}, a very sharp $^{77}$Se nuclear magnetic resonance (NMR) line width \cite{Boehmer2015}, and extremely low level of impurities and defects observed by scanning tunneling microscope topographic images \cite{Kasahara2014,Watashige2015}, all demonstrate that the crystals used in the present study are very clean.  
The tetragonal structure is confirmed by single-crystal X-ray diffraction. The tetragonal [100]$_{\rm T}$/[010]$_{\rm T}$ is along the square edges of the crystals, and below the structural transition, the orthorhombic [100]$_{\rm O}$/[010]$_{\rm O}$ along the diagonal direction.

\section{Magnetization and magnetic torque measurements}

\begin{figure}[t]
\includegraphics[width=0.75\linewidth]{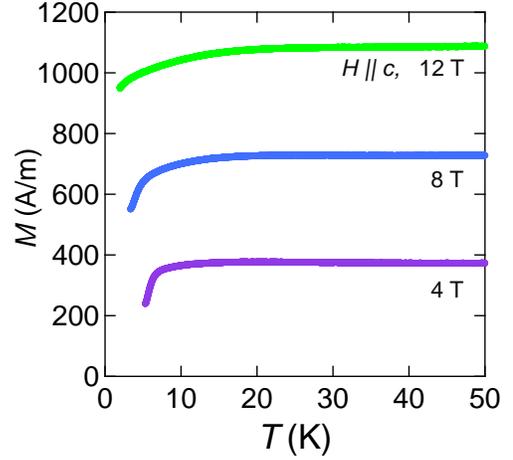}
\caption{Temperature dependence of magnetization for different fields applied along $c$ axis.} 
\end{figure}

\begin{figure}[t]
\includegraphics[width=1.0\linewidth]{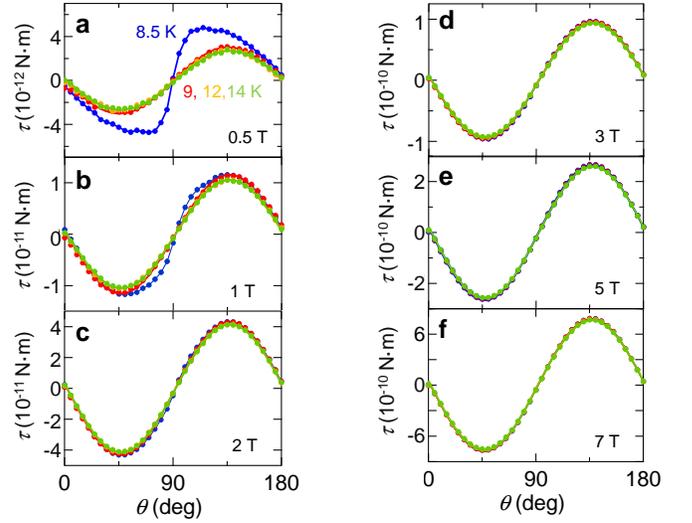}
\caption{Field angle dependence of magnetic torque for various fields and temperatures. $\theta$ is defined the angle between the field and the $c$ axis. } 
\end{figure}

The magnetization was measured using a vibrating sample option (VSM) of the Physical Properties Measurement System (PPMS) by Quantum Design. Figure\,S1 shows temperature dependence of the magnetization in a single crystal of FeSe for several different fields. We obtained the diamagnetic response in the magnetization, $M_{dia}$, by shifting the curves to zero at 30 K, i.e. by subtracting a constant representative of the normal-state magnetization ignoring the small paramagnetic Curie-Weiss contribution.

Magnetic torque is measured by using a micro-cantilever method~\cite{Okazaki2011,Kasahara2012}. 
As illustrated in the inset of Fig.\;2a in the main text, a carefully selected tiny crystal of ideal tetragonal shape with $200 \times 200 \times 5$ $\mu$m$^3$ is mounted on to a piezo-resistive cantilever. Figures\;S2a-f show the magnetic torque $\tau$ measured in various fields, where the field orientation is varied within a plane including the $c$ axis ($\theta=0, 180$ deg) and the field strength $H=|\bm{H}|$ is kept during the rotation. 
The crystals contain orthorhombic domains with typical size of $\sim5\,\mu$m below $T_s$.  
We selected crystals with no detectable ferromagnetic impurities, which show reversible sinusoidal torque curves with a hysteresis component of less than 0.01\% of the total torque (upper inset, Fig.\,2a in the main text).  
At 0.5 and 1\,T (Fig.\;S2a and b), the curves are distorted at $8.5$\,K which is expected in the superconducting state of anisotropic materials \cite{Kogan1988}, while those above 9\,K are perfectly sinusoidal and are described well by $\tau(\theta, H, T) = A\sin(2\theta)$. All the torque curves above 2\,T (Figs.\;S2c-f) are perfectly sinusoidal in the normal state above $T_c$.  
The difference between the $c$-axis and $ab$-plane susceptibilities per unit volume, $\Delta\chi =\chi_c-\chi_{ab}$, is derived from Eq.\;(1) in the main text. We note that the torque signal above $T_c$ is completely reversible without hysteresis, which excludes any ferromagnetic impurity as a source for the enhanced $|\Delta\chi|$ below $T^{\ast}$.

\section{NMR measurements}
\begin{figure}[t]
	\includegraphics[width=0.8\linewidth]{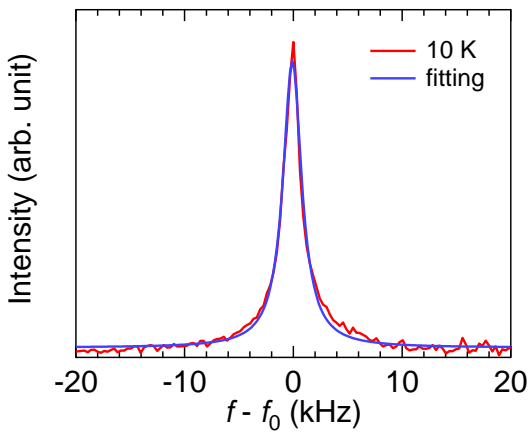}
	\caption{NMR spectrum at 10\,K for $\mu_0H=1$\,T applied along the $c$ axis	(red line). Intensity is plotted against the relative frequency from the resonance frequency at $f_0$. A Lorentzian fit to the data (blue line) shows a narrow FWHM of $\sim2$\,kHz, indicating a homogeneous electronic state.
	}
\end{figure}

$^{77}$Se NMR measurements were performed on a collection of several oriented single crystals, and external fields (1, 2, and 14.5\,T) are applied parallel to the $c$ axis. Since $^{77}$Se has a nuclear spin $I = 1/2$ and thus no electric quadrupole interactions, the resonance linewidth of the NMR spectra are very narrow with full width at half maximum (FWHM) of a couple of kHz (Fig.\;S3). The nuclear spin-lattice relaxation rate $1/T_1$ is evaluated from the recovery curve $R(t)=1-m(t)/m(\infty)$ of the nuclear magnetization $m(t)$, which is the nuclear magnetization at a time $t$ after a saturation pulse. $R(t)$ can be described by $R(t) \propto \exp(-t/T_1)$ with a unique $T_1$ in the whole measured region, indicative of a homogeneous electronic state.
In general, $1/T_1$ for $H \parallel c$ is related to the imaginary part of the dynamical magnetic susceptibility $\chi({\bf q}, \omega)$ by the relation  
\begin{equation}
\frac{1}{T_1T} \propto \sum_{\bf q} A({\bf q})\frac{{\rm Im}\chi({\bf q}, \omega)}{\omega} ,
\end{equation}
where $A({\bf q})$ is the transferred hyperfine coupling tensor along the $c$ axis at the Se site and $\omega= \gamma_n/H$ with $\gamma_n/(2\pi) =8.118$\,MHz/T is the NMR frequency. $1/T_1T$ at the Se site is mainly governed by the magnetic fluctuations at the Fe sites, i.e., particularly in FeSe, the short-lived stripe-antiferromagnetic correlations at ${\bf q} = (\pi, \pi)$ in the tetragonal notation. It should be noted that the superconducting diamagnetism is an orbital effect which is dominated at ${\bf q} =0$ and thus it does not affect the dynamical spin susceptibility at ${\bf q} = (\pi, \pi)$.

\section{Dominant quasiparticle contribution in Nernst signal}

\begin{figure}[b]
\includegraphics[width=1.0\linewidth]{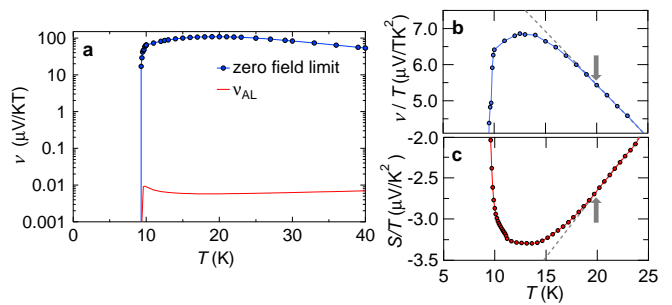}
\caption{{\bf a}, Temperature dependence of the Nernst coefficient $\nu$ in the zero-field limit (blue circles) compared with the expected superconducting fluctuation contribution calculated from the AL theory Eq.\,(S1). {\bf b}, $\nu/T$ as a function of temperature. {\bf c}, Seebeck coefficient divided by temperature $S/T$ as a function of temperature at zero field. Dashed lines are the linear fits at high temperatures. The arrows mark the psudogap temperature $T^*$. } 
\end{figure}

The Nernst signal $N$ is the electric field $E_y$ ($||y$) response to a transverse temperature gradient $\nabla_x T$($||x$) in the presence of a magnetic field $H_z$ ($||z$), and is given by $N \equiv E_y/(-\nabla_x T)$. The Nernst coefficient is defined as $\nu \equiv N/\mu_0H$.

We estimate the Gaussian-type superconducting fluctuation contribution to the Nernst signal by $\nu_{\rm AL}\approx \alpha_{xy}^{\rm AL}\rho_{xx}/(\mu_0 H)$, where $\alpha_{xy}^{\rm AL}$ is the Peltier coefficient \cite{Yamashita2015}. In the Aslamasov-Lakin (AL) theory, this coefficient is given by 
\begin{equation}
\alpha_{xy}^{\rm AL}(T)=\frac{1}{12\pi}\frac{k_{\rm B}e}{\hbar}\frac{\tilde{\xi}_{ab}^2(T)}{\ell_H^2 \tilde{\xi}_c(T)},
\end{equation} 
where $\tilde{\xi}_{i}(T)=\xi_{i}(0)/\sqrt{\ln(T/T_c)}$ $(i=ab$ or $c)$ are the fluctuation coherence lengths parallel to the $ab$ plane and $c$ axis, and $\ell_H=\sqrt{\hbar /2e\mu_0 H}$ is the magnetic length. The red line in Fig.\,S4a represents the AL calculation for the superconducting fluctuation contribution $\nu_{\rm AL}(T)$ by using the resistivity data of the same sample and Eq.\,(S2). Just above $T_c$, the measured data are almost four orders of magnitude larger than the AL estimate. Although the diamagnetic signal is one order of magnitude larger than the AL estimate (Fig.\,2c), both the four-orders-of-magnitude difference and the absence of a diverging behaviour near $T_c$ in $\nu(T)$ demonstrate that in our samples of FeSe the normal quasiparticle contribution dominates over the superconducting fluctuation contribution in the normal state. 

As shown in Figs.\,S4b and S4c,  $\nu/T$ and $S/T$ deviate from the high-temperature linear extrapolation below $T^* \sim 20$\,K.  At lower temperatures, $\nu/T$ and $|S|/T$ both decrease with decreasing $T$ before superconductivity sets in, which is not expected in a Fermi-liquid metal where $\nu/T$ and $S/T$ should become constant in the low-temperature limit.

\section{Field dependence of in-plane resistivity and Nernst signal}

In our clean crystals, a huge magnetoresistance is observed as shown in Fig.\,1a of the main text. Due to the large value of $\omega_c\tau$, where $\omega_c$ is the cyclotron frequency and $\tau$ is the scattering time, $\rho_{xx}$ under strong magnetic fields exhibits a semiconducting behaviour at low temperatures. 
In strong fields, superconducting fluctuations, or excess conductivity, are observed more clearly as they compete with the increase of $\rho_{xx}$.  
To see this, we plot the temperature derivative of the in-plane resistivity, $d\rho_{xx}/dT$, for several magnetic fields (Fig.\,S5a). For each field, $d\rho_{xx}/dT$ exhibits a clear minimum at $T^\ast\sim20$\,K, indicating a slope change due to the emergence of superconducting fluctuations.  

\begin{figure}[b]
\includegraphics[width=1.0\linewidth]{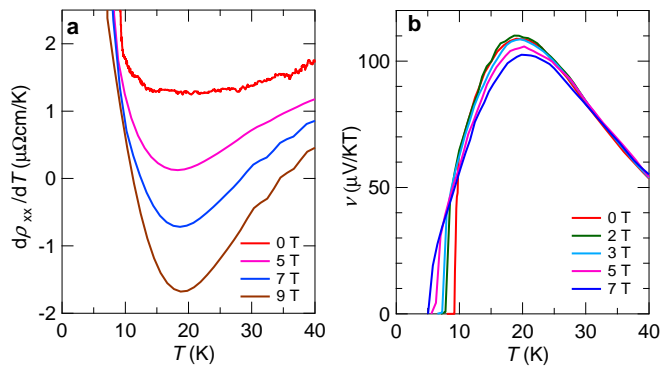}
\caption{{\bf a}, Temperature derivative of the in-plane resistivity $d\rho_{xx}/dT$ and {\bf b}, Nernst signal $\nu$ for different  magnetic fields. } 
\end{figure}

Figure\,S5b shows the temperature dependence of the Nernst coefficient for different fields. 
Similar to $d\rho_{xx}/dT$, $\nu(T)$ at each field exhibits a clear peak at $T^\ast\sim20$\,K, suggesting a change of the energy dependence of the scattering time at the Fermi level. We note that the anomalies in $d\rho_{xx}/dT$ and $\nu(T)$, as well $M_{\rm dia}$ are less sensitive to magnetic field, suggesting that they represent the amplitude fluctuations, while the non-linear behaviour in magnetic torque is readily suppressed by a strong magnetic field. This suggests that the non-linear diamagnetic response in torque is directly related to the phase fluctuations arising from the mode coupling of superconducting fluctuations.

\end{document}